\documentclass[12pt,a4paper]{article}

\newcommand{\la}{\lambda}
\newcommand{\rx}{\mathrm{X}}
\newcommand{\te}{\tau_e}
\newcommand{\xa}{\xi_1}
\newcommand{\xb}{\xi_2}
\newcommand{\xc}{\xi_3}
\newcommand{\ab}{\alpha_2}
\newcommand{\ac}{\alpha_3}
\newcommand{\de}{\Delta}
\newcommand{\oa}{\omega_1}
\newcommand{\ob}{\omega_2}
\newcommand{\xe}{x_{0e}}
\newcommand{\jaa}{J_1^{\mathrm{I}}}
\newcommand{\jab}{J_1^{\mathrm{II}}}
\newcommand{\jac}{J_1^{\mathrm{III}}}
\newcommand{\maa}{m_1^{\mathrm{I}}}
\newcommand{\mab}{m_1^{\mathrm{II}}}
\newcommand{\mac}{m_1^{\mathrm{III}}}
\newcommand{\jba}{J_2^{\mathrm{I}}}
\newcommand{\jbb}{J_2^{\mathrm{II}}}
\newcommand{\jbc}{J_2^{\mathrm{III}}}
\newcommand{\mba}{m_2^{\mathrm{I}}}
\newcommand{\mbb}{m_2^{\mathrm{II}}}
\newcommand{\mbc}{m_2^{\mathrm{III}}}
\newcommand{\pa}{\partial}
\newcommand{\del}{\delta}

\begin{document}

\begin{flushright}
{ }
\end{flushright}
\vspace{1.8cm}

\begin{center}
 \textbf{\Large Extremal Correlator of Three Vertex Operators \\
for Circular Winding Strings in $AdS_5\times S^5$}
\end{center}
\vspace{1.6cm}
\begin{center}
 Shijong Ryang
\end{center}

\begin{center}
\textit{Department of Physics \\ Kyoto Prefectural University of Medicine
\\ Taishogun, Kyoto 603-8334 Japan}
\par
\texttt{ryang@koto.kpu-m.ac.jp}
\end{center}
\vspace{2.8cm}
\begin{abstract}
We study a three-point correlator of the three heavy vertex operators
representing the circular winding string states which are point-like in
$AdS_5$ and rotating with two spins and two winding numbers in $S^5$.
We restrict ourselves to the case that two  of the three 
vertex operators are located at the same point. We evaluate 
semiclassically the specific three-point correlator on a stationary
splitting string trajectory which is mapped to the complex plane 
with three punctures. It becomes an extremal and 4d conformal 
invariant three-point correlator on the boundary. The marginality
condition of the vertex operator is discussed.
\end{abstract}
\vspace{3cm}
\begin{flushleft}
September, 2011 
\end{flushleft}

\newpage
\section{Introduction}

The AdS/CFT correspondence \cite{MGW} has more and more revealed the deep
relations between the $\mathcal{N}=4$ super Yang-Mills (SYM) theory
and the string theory in $AdS_5 \times S^5$. A lot of fascinating results
have been found in the computation of the planar contribution to the 
conformal dimensions of non-BPS operators for any value of the coupling
constant, using integrability \cite{NB}.

From the AdS/CFT correspondence the correlation functions in the 
$\mathcal{N}=4$ SYM theory can be calculated both at weak and at strong
coupling. The three-point correlation functions of BPS operators have been
derived at strong coupling in the supergravity 
approximation \cite{FMM}. There have been in the past 
\cite{AP,GKP,BCF,AT,TS,DY,CIR} various investigations that correlation
functions of operators in $AdS_5 \times S^5$ string theory carrying large
charges of order of string tension $(\sim \sqrt{\la})$ should be 
controled at large $\la$ by the semiclassical string solutions.

From the semiclassical procedure the two-point correlators of heavy string
vertex operators have been constructed \cite{EB,JSW,BT} where  relevant
string surfaces ending at the AdS boundary saturate the correlators.
There has been  an extension to certain three-point correlator of two
heavy string vertex operators with large charges and one light operator 
representing a BPS state or a non-BPS string light mode \cite{ZA,CMS,RT},
where such correlator can be computed by evaluating the light operator 
on the same classical string solution as saturates the corresponding
two-point correlator of the heavy string vertex operators.
Further constructions of the three-point correlators of two various 
heavy string states and a certain light mode have been performed 
\cite{RH,SR,GG,JRT,PL}. This approach has been extended to the four-point
correlators of two heavy vertex operators and two light operators
\cite{EBT}, giant gravitons \cite{BCW}, finite-size
effects \cite{AB} and various other aspects \cite{BLP}.  
The correlators involving
Wilson loops and a light local operator have been investigated \cite{LAT}.

In the $\mathcal{N}=4$ SYM theory side the planar three-point correlators
of single trace gauge-invariant operators have been studied at
weak coupling \cite{OT}. Using the integrability techniques the
three-point correlators involving two operators in the SU(3) sector
have been computed at weak coupling \cite{EGS,EGV}, where a precise match
between weak and strong coupling is observed for the correlators of two
non-BPS operators in the Frolov-Tseytlin limit and one short BPS
operator \cite{EGV}. This matching has been further demonstrated for
the three-point correlators involving two operators in the SL(2,R) closed
subsector of $\mathcal{N}=4$ SYM theory \cite{GE} and for the four-point
correlators \cite{CE}.

The three-point correlator of three BMN string states which are point-like
in $AdS_5$ and rotating along a great circle of $S^5$ has been computed by
making a simple assumption that the three cylinders associated with three
external states may be joined
together at an intersection point. By taking advantage of the conformal 
symmetry of $AdS_5$ instead of extremizing with respect to the
intersection point the 4d conformal invariant dependence on the 
three positions of the operator insertion points has been 
extracted \cite{JSW}. In ref. \cite{KM} using a light-cone gauge for the
worldsheet theory the three-point correlator of BMN vertex operators
has been computed by explicitly minimizing the action upon varying
the intersection point of three Euclidean BMN strings. Further
the three-point correlator of  the three heavy string vertex operators
representing the circular spinning strings with winding numbers in $S^5$
has been constructed to have a 4d conformal invariant expression
by solving the minimization problem for the intersection point.

The classical splitting of strings has been extensively studied \cite{VR},
mostly in flat Minkowski space. In the context of the AdS/CFT 
correspondence, the splitting of the spinning string with large spin
in $R_t \times S^5$ has been investigated \cite{PPZ}, and the splitting
of the folded spinning string in $AdS_3$ together with spins and winding
numbers in $S^3$ has been studied \cite{EM}, 
where a special splitting string
solution describing the evolution of the final states after the string
splitting is presented. Using the classical integrability of the string
$\sigma$-model the general classical splitting string solution in
$R_t \times S^3$ has been constructed \cite{BV}. 

Using the vertex operator prescription we will compute a special type of
three-point correlator of the three heavy string vertex operators
where two locations of the three vertex operators 
coincide and each vertex operator
represents the circular string state that is  point-like in $AdS_5$ and
rotating with two spins and two winding numbers in $S^5$.
By taking account of the splitting process of a circular winding string
into two strings we will construct a stationary splitting string
trajectory which controls the three-point correlator.
Through the stationary surface sourced by the three vertex operators
the semiclassically evaluated correlator will be 
expressed as an extremal and 4d conformal invariant
three-point correlator in the boundary theory. It will be also regarded 
as a 2d conformal invariant three-point correlator where the three points
describe the locations of three punctures that are the vertex operator
positions on the worldsheet. 

\section{Three-point correlator and string splitting}

Based on the vertex operator prescription \cite{AT,EB,BT} we consider
a three-point correlation function of the three heavy string vertex 
operators which are associated with the circular spinning string solution
describing a point-like string in $AdS_5$ and a rotating string
with spins and winding numbers $(J_1, m_1)$ and $(J_2, m_2)$ in $S^5$. 

The embedding coordinates $Y_M \; (M=0, \cdots,5)$ for the Minkowski 
signature $AdS_5$ are expressed in terms of the global coordinates
$(t, \rho, \theta, \phi_1, \phi_2)$ as
\begin{eqnarray}
Y_5 + iY_0 &=& \cosh \rho e^{it}, \;\; Y_1 + iY_2 = \sinh \rho 
\cos \theta e^{i\phi_1}, \;\;Y_3 + iY_4 = \sinh \rho \sin \theta
e^{i\phi_2}, \nonumber \\
Y^MY_M &=& -Y_5^2 + Y^mY_m + Y_4^2 = -1, \hspace{1cm} 
Y^mY_m = -Y_0^2 + Y_iY_i
\end{eqnarray}
with $m=0,1,2,3, i=1,2,3$. These coordinates are related with the
Poincare coordinates $(z, x^m), ds^2 = z^{-2}(dz^2 + dx^mdx_m)$ 
\begin{equation}
Y_m = \frac{x_m}{z}, \hspace{1cm} Y_4 = \frac{1}{2z}(-1 + z^2 + x^mx_m),
\hspace{1cm} Y_5 = \frac{1}{2z}(1 + z^2 + x^mx_m)
\end{equation}
with $x^mx_m = -x_0^2 + x_i^2$. For $S^5$ the embedding coordinates are 
also defined by 
\begin{eqnarray}
\rx_1 &\equiv& X_1 + iX_2 = \sin\gamma \cos\psi e^{i\varphi_1} =
r_1e^{i\varphi_1}, \hspace{1cm}
\rx_2 \equiv X_3 + iX_4 = \sin\gamma \sin\psi e^{i\varphi_2}= 
r_2e^{i\varphi_2}, \nonumber \\
\rx_3 &\equiv& X_5 + iX_6 = \cos\gamma e^{i\varphi_3}= 
r_3e^{i\varphi_3}, \hspace{1cm} \sum_{k=1}^3 r_k^2 = 1.
\end{eqnarray}

Both the worldsheet time $\tau$ and the global AdS time $t$ 
are rotated to the Euclidean ones simultaneously 
\begin{equation}
\te = i\tau, \hspace{1cm} t_e = it,
\label{eu}\end{equation}
which lead to the similar rotations for the time-like coordinates
\begin{equation}
Y_{0e} = iY_0, \hspace{1cm} \xe = ix_0.
\end{equation}

We perform the following conformal  transformation to map the Euclidean 
$(\te, \sigma)$ worldsheet into the $\xi$ complex plane  with
three punctures located at $ \xi = \xa \;(\te = -\infty), \; 
\xi = \xb \;(\te = \infty), \; \xi = \xc \;(\te = \infty)$
\begin{equation}
e^{\te + i\sigma} = \frac{\xi - \xa}{(\xi - \xb)^{\ab}(\xi - \xc)^{\ac}},
\hspace{1cm} \ab >0,\; \ac>0.
\label{con}\end{equation}
We choose
\begin{equation}
\ab + \ac =1
\label{sum}\end{equation}
so that the infinite point $\xi = \infty$ is mapped to $(\te, \sigma) = 
(0,0)$. These three arbitrary finite positions $\xa, \xb, \xc$ will be 
regarded as the positions where three vertex operators are 
inserted on the string surface. 

Let us consider the vertex operator of dimension $\de$ which describes 
a point-like string located at the origin in $AdS_5$ and the 
circular spinning string with quantum numbers like spins $(J_1, J_2)$
and winding numbers $(m_1, m_2)$ in $S^5$ \cite{ART}
\begin{equation}
t = \kappa \tau, \;\; \rho = 0, \;\; \theta = 0, \;\;
\phi_1 = 0, \;\;\varphi_1 = \oa\tau - m_1\sigma, \;\; \varphi_2 = 
\ob\tau + m_2\sigma, \;\; \gamma = \frac{\pi}{2}.
\label{sol}\end{equation}
For the integrated vertex operator 
$V(x') = \int d^2\xi V( x(\xi) - x', \cdots)$ specified by the four 
coordinates $x'_m=(\xe', x_i')$ on the boundary of the
$AdS_5$ space, we define
\begin{equation}
V(a) =  \int d^2\xi [ z + z^{-1}(\xe  - a )^2]^{-\de}
(X_1 + iX_2)^{J_1}(X_3 + iX_4)^{J_2},
\end{equation}
where the location of the vertex operator in the boundary is chosen by
$x_m'=(a,0,0,0)$ and the winding number dependences are implicitly
included through the angular coordinates.

We use this vertex operator to compute a special type of correlator of 
three vertex operators
\begin{equation}
< V_{\de^{\mathrm{I}},\jaa,\maa,\jba,\mba}(-a) V_{\de^{\mathrm{II}}
,\jab,\mab,\jbb,\mbb}^*(a) 
V_{\de^{\mathrm{III}},\jac,\mac,\jbc,\mbc}^*(a) >,
\label{cov}\end{equation}
where we restrict ourselves to the case that the second and the third 
vertex operators are located at the same point on the boundary.
The Euclidean continuation (\ref{eu}) of (\ref{sol}) is given by
\begin{equation}
t_e = \kappa\te, \;\; \rho = 0, \;\; \varphi_1 = -i\oa\te - m_1\sigma,
\;\; \varphi_2 = -i\ob\te + m_2\sigma.
\label{eso}\end{equation} 
For the Euclidean point-like string in $AdS_5$ we make a dilatation
such that $\xe(\infty) = a, \xe(-\infty) = -a$ at the boundary we have
\begin{equation}
z = \frac{a}{\cosh \kappa\te}, \hspace{1cm} \xe = a \tanh \kappa\te
\label{zx}\end{equation}
and the Euclidean spinning string configuration in $S^5$ is specified by
\begin{equation}
\rx_1(\sigma, \te) = r_1 e^{\oa\te - im_1\sigma}, \hspace{1cm}
\rx_2(\sigma, \te) = r_2 e^{\ob\te + im_2\sigma}
\label{xr}\end{equation}
with $r_1 = \cos \psi, r_2 = \sin \psi, \psi = \mathrm{const}$.

Here we consider a splitting of the classical circular 
spinning string with winding numbers in $S^5$. Under the 
Schwarz-Christoffel transformation (\ref{con})
\begin{equation}
\rho \equiv \te + i\sigma = \ln (\xi - \xa) - \ab \ln(\xi - \xb)
- \ac \ln(\xi - \xc),
\label{sc}\end{equation}
the ends of string worldsheet at the initial 
and final times $\te = -\infty$
and $\te = \infty$ in the $\rho$ complex  plane are transformed onto
the three points $\xa$ and $\xb, \xc$ in the $\xi$ complex plane.
The mapping (\ref{sc}) shows that an initial incoming closed string I
splits into an outgoing closed string II with length $2\pi \ab$ and
an outgoing closed string III with length $2\pi\ac$.
Although we are working in a Euclidean formulation, we refer to each 
string state as ``incoming" and ``outgoing". 

The worldsheet in the $\rho$ complex plane 
is expressed as the `pair of pants'
diagram and consists of three cylinders, which are parameterized by
\begin{eqnarray}
W_{\mathrm{I}} &=& \{ (\sigma,\te) |\; 0< \sigma \le 2\pi, \; 
\te < \te^0 \}, \nonumber \\
W_{\mathrm{II}} &=& \{ (\sigma,\te) |\;0< \sigma \le 2\pi\ab, \; 
\te > \te^0 \}, \nonumber \\  
W_{\mathrm{III}} &=& \{ (\sigma,\te) |\; 2\pi\ab< \sigma \le 2\pi, 
\; \te >  \te^0 \}, 
\label{thw}\end{eqnarray}
where $0 < \ab < 1$ and the $\sigma$-interval is periodically
identified in each case. At the local interaction point specified by
$\sigma^0 \equiv 2\pi\ab, \te^0$ the initial closed string I breaks into
two fragments II and III.  

From (\ref{xr}) we express the initial circular spinning string solution
$\rx_k^{\mathrm{I}}(\sigma,\te), k =1,2$ as
\begin{equation}
\rx_k^{\mathrm{I}}(\sigma,\te) = \{ \rx_k(\sigma,\te), \;\;\mathrm{for}
\;\; \te \le \te^0, \; 0 < \sigma \le 2\pi \}, \; m_k^{\mathrm{I}} = m_k.
\label{xon}\end{equation}
We assume that at the time $\te = \te^0$, the two points 
$\sigma = 0, \sigma = 2\pi\ab$ on the initial string coincide 
in target space and their velocities agree
\begin{equation}
\rx_k^{\mathrm{I}}(0,\te^0) = \rx_k^{\mathrm{I}}(2\pi\ab,\te^0), 
\;\; \pa_{\te}\rx_k^{\mathrm{I}}(0,\te^0) = 
\pa_{\te}\rx_k^{\mathrm{I}}(2\pi\ab,\te^0), 
\label{cos}\end{equation}
that is the consistency condition of the self-interaction 
for the closed string splitting to be possible.
The continuity of the string variables  at the interaction time $\te^0$
is imposed as
\begin{eqnarray}
\left. \begin{array}{c} 
\rx_k^{\mathrm{I}}(\sigma, \te^0) =  \rx_k^{\mathrm{II}}(\sigma, \te^0) \\
\pa_{\te}\rx_k^{\mathrm{I}}(\sigma, \te^0) = \pa_{\te}
\rx_k^{\mathrm{II}}(\sigma, \te^0) \end{array} \right\} \;\;\mathrm{for} 
\;\;0 < \sigma  \le 2\pi\ab, \nonumber \\
\left. \begin{array}{c} \rx_k^{\mathrm{I}}(\sigma, \te^0) =  
\rx_k^{\mathrm{III}}(\sigma, \te^0) \\ \pa_{\te}\rx_k^{\mathrm{I}}(\sigma,
\te^0) = \pa_{\te}\rx_k^{\mathrm{III}}(\sigma, \te^0)
\end{array} \right\} \;\;\mathrm{for}\;\; 2\pi\ab < \sigma  \le 2\pi.
\label{cty}\end{eqnarray}
The two outgoing closed string solutions are required to satisfy 
new periodicity conditions
\begin{equation}
\rx_k^{\mathrm{II}}(\sigma + 2\pi\ab,\te) = \rx_k^{\mathrm{II}}
(\sigma,\te), \;\;
\rx_k^{\mathrm{III}}(\sigma + 2\pi\ac,\te) = \rx_k^{\mathrm{III}}
(\sigma,\te).
\label{per}\end{equation}

Now we take a simple separation that the initial circular spinning string
solution $\rx_k^{\mathrm{I}}(\sigma,\te)$
remains valid in the outgoing two regions $W_{\mathrm{II}}, 
W_{\mathrm{III}}$ 
\begin{eqnarray}
\rx_k^{\mathrm{II}}(\sigma,\te) &=& \{ \rx_k(\sigma,\te), \;\;\mathrm{for}
\;\; \te^0 \le \te, \; 0 < \sigma \le 2\pi\ab \}, \nonumber \\
\rx_k^{\mathrm{III}}(\sigma,\te) &=& \{ \rx_k(\sigma,\te), \;\;
\mathrm{for}\;\; \te^0 \le \te, \; 2\pi\ab < \sigma \le 2\pi \}.
\label{xtw}\end{eqnarray}
We regard (\ref{xtw}) as simple special solutions for the outgoing strings
which obey the boundary condition (\ref{cty}) and the relevant equations 
of motion. 

The consistency condition (\ref{cos}) implies 
$m_k^{\mathrm{I}}\ab \in Z^+$ so that we put
\begin{equation}
m_k^{\mathrm{II}} = m_k^{\mathrm{I}}\ab, \;\;(k = 1, 2)
\label{mtw}\end{equation}
by introducing positive integers $m_k^{\mathrm{II}}$. The interaction 
position is fixed as $\sigma^0 = 2\pi m_k^{\mathrm{II}}/m_k^{\mathrm{I}}$.
The periodicity condition (\ref{per}) constrains the behaviors of
the two fragments to yield
\begin{equation}
m_k^{\mathrm{I}}\ab \in Z^+, \hspace{1cm} m_k^{\mathrm{I}}\ac = 
m_k^{\mathrm{I}}( 1 - \ab)\in Z^+,
\end{equation}
which hold owing to (\ref{mtw}). We introduce positive integers
$m_k^{\mathrm{III}}$ again to put
\begin{equation}
m_k^{\mathrm{III}} = m_k^{\mathrm{I}}\ac, \;\; (k =1, 2).
\label{mth}\end{equation}

After rescaling of the worldsheet space coordinate $\sigma$ to
$\sigma^{\mathrm{II}}, \sigma^{\mathrm{III}}$ in such a way that 
$0 < \sigma \le 2\pi\ab  \rightarrow 0 < \sigma^{\mathrm{II}} \le 2\pi$, 
and $2\pi\ab  < \sigma \le 2\pi  \rightarrow 0 < 
\sigma^{\mathrm{III}} \le 2\pi$
\begin{equation}
\sigma^{\mathrm{II}} = \frac{\sigma}{\ab}, \hspace{1cm} 
\sigma^{\mathrm{III}} = \frac{\sigma - 2\pi\ab}{\ac}, 
\end{equation}
we rewrite the two outgoing string configurations (\ref{xtw}) as
\begin{eqnarray}
\rx_k^{\mathrm{II}}(\sigma^{\mathrm{II}}, \te) &=& r_k e^{\omega_k\te
 + (-1)^kim_k^{\mathrm{II}} \sigma^{\mathrm{II}}},  \nonumber \\
\rx_k^{\mathrm{III}}(\sigma^{\mathrm{III}}, \te) &=& r_k e^{\omega_k\te 
+ (-1)^kim_k^{\mathrm{III}} \sigma^{\mathrm{III}}},
\label{tou}\end{eqnarray}
which are specified by the common frequencies $\omega_k$.
From these expressions the positive integers $m_k^{\mathrm{II}}$ 
are interpreted as the winding numbers of the string II on the outgoing
cylinder $W_{\mathrm{II}}$, while the positive integers 
$m_k^{\mathrm{III}}$ as the winding numbers of the string III on the 
outgoing cylinder $W_{\mathrm{III}}$. We observe that in this simple 
separation of string the winding numbers are conserved 
owing to (\ref{sum})
\begin{equation}
m_k^{\mathrm{I}} = m_k^{\mathrm{II}} + m_k^{\mathrm{III}}, 
\;\; (k = 1, 2).
\end{equation}

Thus the splitting process is characterized by the separations of the
winding numbers of the initial incoming string through the
self-interaction of the closed string, where the winding number
$\maa$ is separated into $\mab$ and $\mac$ in the same way as
$\mba$ into $\mbb$ and $\mbc$. The world surface does not change such that
under suitable rescaling of the worldsheet space coordinate $\sigma$,
the two outgoing string solutions (\ref{tou}) 
described by the corresponding 
winding numbers become the same type expression as the initial 
string solution.

The energy of the initial string is given by
\begin{equation}
E^{\mathrm{I}} = \sqrt{\la}\int_0^{2\pi} \frac{d\sigma}{2\pi} \kappa = 
\kappa \sqrt{\la}
\label{eon}\end{equation}
and the energies of the two fragments are obtained by
\begin{eqnarray}
E^{\mathrm{II}} &=& \sqrt{\la}\int_0^{2\pi\ab} \frac{d\sigma}{2\pi} \kappa
 = \ab \kappa \sqrt{\la}, \nonumber \\
E^{\mathrm{III}} &=& \sqrt{\la}\int_{2\pi\ab}^{2\pi} \frac{d\sigma}{2\pi}
\kappa = \ac \kappa \sqrt{\la}.
\label{etw}\end{eqnarray}
Owing to (\ref{sum}) the string energy is conserved 
in the splitting process 
\begin{equation}
E^{\mathrm{I}} = E^{\mathrm{II}} + E^{\mathrm{III}}.
\label{coe}\end{equation}
Similarly the two spins $\jaa, \jba$ of the initial string are given by
\begin{equation}
\jaa = \sqrt{\la}\int_0^{2\pi} \frac{d\sigma}{2\pi}
r_1^2\oa = \sqrt{\la}r_1^2\oa, \;\;
\jba = \sqrt{\la}\int_0^{2\pi} \frac{d\sigma}{2\pi}
r_2^2\ob = \sqrt{\la}r_2^2\ob,
\label{jon}\end{equation}
while the two spins  $\jab, \jbb$ of the fragment II and
the two spins  $\jac, \jbc$ of the fragment III are expressed as
\begin{eqnarray}
\jab &=& \ab\sqrt{\la}r_1^2\oa, \hspace{1cm} \jbb = \ab\sqrt{\la}r_2^2\ob,
\nonumber \\
\jac &=& \ac\sqrt{\la}r_1^2\oa, \hspace{1cm} \jbc = \ac\sqrt{\la}r_2^2\ob.
\label{jtw}\end{eqnarray}
We have the conservation of each spin
\begin{equation}
\jaa = \jab + \jac, \hspace{1cm} \jba = \jbb + \jbc.
\end{equation}

\section{Extremal three-point correlator}

In order to compute semiclassically the correlator 
(\ref{cov}), we express the Euclidean action accompanied with the vertex
contributions as an integral over the $\xi$ complex  plane
with three punctures describing the splitting string worldsheet 
\begin{eqnarray}
A_e &=& \frac{\sqrt{\la}}{\pi}\int d^2\xi \left[ \frac{1}{z^2}
( \pa z\bar{\pa}z + \pa \xe \bar{\pa}\xe  ) + 
\frac{1}{2}\sum_{k=1}^2 (\pa \rx_k \bar{\pa}\bar{\rx}_k +
\bar{\pa} \rx_k \pa\bar{\rx}_k ) \right] \nonumber \\
 &-& \de^{\mathrm{I}} \int d^2\xi  \del^2(\xi - \xa) \ln 
\frac{z}{z^2 + (\xe + a)^2} - \de^{\mathrm{II}} \int d^2\xi 
\del^2(\xi - \xb) \ln \frac{z}{z^2 + (\xe - a)^2}
\nonumber \\
&-& \de^{\mathrm{III}} \int d^2\xi  \del^2(\xi - \xc) \ln \frac{z}{z^2
+ (\xe - a)^2} - \sum_{k=1}^2 J_k^{\mathrm{I}} \int d^2\xi 
\del^2(\xi - \xa) \ln r_ke^{i\varphi_k} \nonumber \\
 &-& \sum_{k=1}^2 J_k^{\mathrm{II}} \int d^2\xi \del^2(\xi - \xb) \ln 
r_ke^{-i\varphi_k} - \sum_{k=1}^2 J_k^{\mathrm{III}} \int d^2\xi 
\del^2(\xi - \xc) \ln r_ke^{-i\varphi_k}.
\label{act}\end{eqnarray}

We will show that the Euclidean string solution (\ref{zx}), (\ref{xr})
expressed in terms of the complex worldsheet coordinate $\xi$ becomes the
stationary splitting string trajectory in the presence of three vertex
operators as source terms. The equation of motion for $\xe$ is given by
\begin{eqnarray}
\pa \frac{\bar{\pa}\xe}{z^2} + \bar{\pa} \frac{\pa\xe}{z^2} &=&
\frac{2\pi}{\sqrt{\la}} \biggl[ \de^{\mathrm{I}}
\frac{\xe + a}{z^2 + (\xe + a)^2}
\del^2(\xi - \xa) + \de^{\mathrm{II}}\frac{\xe - a}{z^2 + (\xe - a)^2}
\del^2(\xi - \xb) \nonumber \\
&+& \de^{\mathrm{III}}\frac{\xe - a}{z^2 + (\xe - a)^2}
\del^2(\xi - \xc)\biggr].
\label{eqx}\end{eqnarray}
The inversion of the conformal transformation (\ref{con}) is expressed as
\begin{eqnarray}
\te &=& \frac{1}{2}\ln \frac{(\xi - \xa)(\bar{\xi}- \bar{\xa})}
{(\xi - \xb)^{\ab}(\bar{\xi} - \bar{\xb})^{\ab} 
(\xi - \xc)^{\ac}(\bar{\xi} - \bar{\xc})^{\ac}}, \nonumber \\
\sigma &=& \frac{1}{2i}\ln \frac{(\xi - \xa)(\bar{\xi} - \bar{\xb})^{\ab}
(\bar{\xi} - \bar{\xc})^{\ac} }
{(\bar{\xi}- \bar{\xa})(\xi - \xb)^{\ab}(\xi - \xc)^{\ac} },
\end{eqnarray}
which produce
\begin{equation}
(\pa\bar{\pa} + \bar{\pa}\pa)\te = \pi [ \del^2(\xi - \xa)  - 
\ab\del^2(\xi - \xb) - \ac\del^2(\xi - \xc) ], \hspace{1cm}
(\pa\bar{\pa} + \bar{\pa}\pa)\sigma = 0.
\label{pat}\end{equation}

The substitution of (\ref{zx}) into (\ref{eqx}) leads to 
\begin{equation}
\kappa (\pa\bar{\pa} + \bar{\pa}\pa)\te = \frac{\pi}{\sqrt{\la}}
[ \de^{\mathrm{I}} \del^2(\xi - \xa) - \de^{\mathrm{II}}\del^2(\xi - \xb)
 - \de^{\mathrm{III}}\del^2(\xi - \xc) ],
\label{ka}\end{equation}
which is satisfied through (\ref{pat}) if each dimension is given by
\begin{equation}
\de^{\mathrm{I}} = \kappa \sqrt{\la}, \;\; \de^{\mathrm{II}} = \ab\kappa
\sqrt{\la}, \;\; \de^{\mathrm{III}} = \ac\kappa \sqrt{\la},
\label{de}\end{equation}
which correspond to (\ref{eon}), (\ref{etw}).

We turn to the equation of motion for $z$ 
\begin{eqnarray}
\pa \frac{\bar{\pa}z}{z^2} + \bar{\pa}  
\frac{\pa z}{z^2}  + \frac{2}{z^3}(\pa z \bar{\pa}z + 
\pa \xe \bar{\pa}\xe) =  \frac{\pi}{\sqrt{\la}z}\biggl[ 
\de^{\mathrm{I}}\frac{z^2 - (\xe + a)^2}{z^2 + (\xe + a)^2}\del^2
(\xi - \xa) \nonumber \\
+ \de^{\mathrm{II}}\frac{z^2 - (\xe - a)^2}{z^2 + (\xe - a)^2}\del^2
(\xi - \xb)+ \de^{\mathrm{III}}\frac{z^2 - (\xe - a)^2}{z^2 + (\xe - a)^2}
\del^2(\xi - \xc)\biggr].
\end{eqnarray}
Since the limits $\te \rightarrow -\infty$ and $\te \rightarrow \infty$
correspond to the limits $\xi \rightarrow \xa$ and $\xi 
\rightarrow \xi_{2,3}$ respectively it becomes 
\begin{equation}
\kappa (\pa\bar{\pa} + \bar{\pa}\pa)\te = -\frac{\pi}
{\sqrt{\la}\tanh \kappa\te}
[ \de^{\mathrm{I}} \del^2(\xi - \xa) + \de^{\mathrm{II}}\del^2(\xi - \xb)
 + \de^{\mathrm{III}}\del^2(\xi - \xc) ],
\end{equation}
which is reduced to (\ref{ka}).

The kinetic term for $S^5$ in (\ref{act}) is rewritten by
$\sqrt{\la}/\pi \int d^2\xi \sum_k( \pa r_k\bar{\pa}r_k + r_k^2
\pa \varphi_k\bar{\pa}\varphi_k )$. Therefore the equations of motion for
$\varphi_k \; (k=1,2)$ read
\begin{equation}
r_k^2 (\pa\bar{\pa} + \bar{\pa}\pa)\varphi_k = \frac{i\pi}{\sqrt{\la}}
[ J_k^{\mathrm{I}} \del^2(\xi - \xa) - J_k^{\mathrm{II}}\del^2(\xi - \xb)
 - J_k^{\mathrm{III}}\del^2(\xi - \xc) ],
\end{equation}
which are satisfied by $\varphi_k$ in (\ref{eso}) 
through (\ref{jon}), (\ref{jtw}).
The equations of motion for $r_k$ have additional contributions to the 
singular part. Here we assume that they are still satisfied 
by constant $r_k$. 

Now the three-point correlator can be calculated semiclassically by
evaluating string action with source terms on the stationary splitting 
string trajectory. It is convenient to go back to the Euclidean 
cylindrical worldsheet coordinates $(\te,\sigma)$
through (\ref{con}) for computing the string action in (\ref{act})
\begin{eqnarray}
A_{str} &=& \frac{\sqrt{\la}}{4\pi} \left[ \int_{\te(\xa)}^{\te^0}d\te
\int_0^{2\pi} d\sigma + \int_{\te^0}^{\te(\xb)}d\te
\int_0^{2\pi\ab} d\sigma + \int_{\te^0}^{\te(\xc)}d\te
\int_{2\pi\ab}^{2\pi} d\sigma \right] L_{str}, \nonumber \\ 
L_{str} &=& \frac{1}{z^2}( (\pa_{\te}z)^2 +  (\pa_{\te}\xe)^2 ) +
\sum_{k=1}^2r_k^2( (\pa_{\te}\varphi_k)^2 + (\pa_{\sigma}\varphi_k)^2 )
\label{ast}\end{eqnarray}
with
\begin{equation}
\tau_{e}(\xi_k) = \frac{1}{2}( \ln |\xi - \xa |^2 - \ab\ln |\xi - \xb |^2
- \ac\ln |\xi - \xc |^2 )|_{\xi \rightarrow \xi_k},
\end{equation}
where the integral region is divided into three sectors according to 
(\ref{thw}). 

Taking into account that the two outgoing 
string solutions are identical to the
incoming string solution we substitute the stationary string solution
(\ref{eso}) or the equivalent one (\ref{zx}) with (\ref{xon}), (\ref{xtw})
into (\ref{ast}) and then perform the integral over $\te$ and $\sigma$ 
to have
\begin{equation}
A_{str} = \frac{\sqrt{\la}}{2} [ \kappa^2 +  
\sum_{k=1}^2r_k^2 ( -\omega_k^2 + (m_k^I)^2 ) ]( \ab\ln|\xa - \xb|^2 
- \ab\ac\ln|\xb - \xc|^2 + \ac\ln|\xc - \xa|^2  ),
\end{equation}
where the one-point function divergence is ignored.
It is observed that the dependence on $\te^0$ disappears, although
the splitting time $\te^0$ is specified by $\ab, \xi_i (i=1,2,3)$
through the turning point condition.
The source terms associated with $AdS_5$ in (\ref{act}) are evaluated
using the delta-function as 
\begin{eqnarray}
A_{sour}(AdS_5) &=& -\de^{\mathrm{I}}\left[ \frac{\kappa}{2}( \ab\ln|\xa 
- \xb|^2  + \ac\ln|\xc - \xa|^2) - \ln 2a \right]  \nonumber \\
&-&\de^{\mathrm{II}}\left[ \frac{\kappa}{2}( \ln|\xa - \xb|^2  - 
\ac\ln|\xb - \xc|^2)  - \ln 2a \right]  \nonumber \\
&-&\de^{\mathrm{III}}\left[ \frac{\kappa}{2}( \ln|\xc - \xa|^2  - 
\ab\ln|\xb - \xc|^2) - \ln 2a \right] .
\end{eqnarray}
The source terms associated with $S^5$ in (\ref{act}) read
\begin{eqnarray}
&A_{sour}(S^5) =  \frac{1}{2}\ln |\xa - \xb|^2[ \oa\jab + \ob\jbb + 
\ab(\oa\jaa + \ob\jba) ]& \nonumber \\
&- \frac{1}{2}\ln |\xb - \xc|^2[ \ab(\oa\jac + \ob\jbc) +
\ac(\oa\jab + \ob\jbb) ]& \nonumber \\ 
&+ \frac{1}{2}\ln |\xc - \xa|^2[ \ac(\oa\jaa + \ob\jba) +
\oa\jac + \ob\jbc ]& \nonumber \\ 
&+ \frac{1}{2}(\maa\jaa - \mba\jba)\ln\frac{(\bar{\xa} - \bar{\xb})^{\ab}
(\bar{\xa} - \bar{\xc})^{\ac}}{(\xa - \xb)^{\ab}(\xa - \xc)^{\ac}}
- \frac{1}{2}(\maa\jab - \mba\jbb)\ln\frac{(\xb - \xa)
(\bar{\xb} - \bar{\xc})^{\ac}}
{ (\bar{\xb} - \bar{\xa}) (\xb - \xc)^{\ac}}&
\nonumber \\
&- \frac{1}{2}(\maa\jac - \mba\jbc)\ln\frac{(\xc - \xa)
(\bar{\xc} - \bar{\xb})^{\ab}}
{ (\bar{\xc} - \bar{\xa})(\xc - \xb)^{\ab}}.&
\label{aso}\end{eqnarray}

Since the coefficients in the last two terms in (\ref{aso}) are 
expressed as
\begin{equation}
\maa\jab - \mba\jbb = \ab(\maa\jaa - \mba\jba), \;\;
\maa\jac - \mba\jbc = \ac(\maa\jaa - \mba\jba),
\end{equation}
we put
\begin{equation}
\maa\jaa = \mba\jba
\label{mj}\end{equation}
to obtain a 2d conformal invariant expression
through (\ref{jon}) and (\ref{jtw})
\begin{eqnarray}
&<& V_{\de^{\mathrm{I}},\jaa,\maa,\jba,\mba}(-a) V_{\de^{\mathrm{II}}
,\jab,\mab,\jbb,\mbb}^*(a) V_{\de^{\mathrm{III}},\jac,\mac,\jbc,\mbc}^*(a)
 > \nonumber \\
&\approx& \int d^2\xa d^2\xb d^2\xc e^{-A_{str} - A_{sour}(AdS_5)
- A_{sour}(S^5)} \nonumber \\
&\approx& \frac{1}{(2a)^{\de^{\mathrm{I}} + \de^{\mathrm{II}} + 
\de^{\mathrm{III}} }}
\int d^2\xa d^2\xb d^2\xc
\frac{1}{|\xa - \xb|^{ \sqrt{\la}[ \ab( \kappa^2 + \sum_k r_k^2(\omega_k^2
+ (m_k^I)^2 ) -\frac{\kappa}{\sqrt{\la}}(\ab \de^{\mathrm{I}} + 
\de^{\mathrm{II}})] }  } \nonumber \\
&\times& \frac{1}{|\xb - \xc|^{-\sqrt{\la}[ \ab\ac
( \kappa^2 + \sum_k r_k^2(\omega_k^2 + (m_k^I)^2 ) -\frac{\kappa}
{\sqrt{\la}}(\ac \de^{\mathrm{II}} + \ab\de^{\mathrm{III}})] }  } 
\nonumber \\ 
&\times& \frac{1}{|\xc - \xa|^{\sqrt{\la}[ \ac
( \kappa^2 + \sum_k r_k^2(\omega_k^2 + (m_k^I)^2 ) -\frac{\kappa}
{\sqrt{\la}}(\ac \de^{\mathrm{I}} + \de^{\mathrm{III}})] }  }.
\label{cor}\end{eqnarray}
This correlator is regarded as a space-time three-point correlator
of the boundary theory at strong coupling where two space-time points
coincide, while it is in itself defined as
 a worldsheet three-point correlator
of the string theory where the three points are associated with 
the three punctures on the worldsheet.
The dimensions $\de^{\mathrm{I}}, \de^{\mathrm{II}}, \de^{\mathrm{III}}$
 are identified with the incoming string energy $E^{\mathrm{I}}$ and 
the outgoing string energies 
$E^{\mathrm{II}}, E^{\mathrm{III}}$ respectively.

Owing to the relations $\de^{\mathrm{II}} = \ab \de^{\mathrm{I}}, 
\de^{\mathrm{III}}  = \ac \de^{\mathrm{I}}$ in (\ref{de}) the 
three-point correlator (\ref{cor}) turns out to be
\begin{equation}
\frac{1}{(2a)^{\de^{\mathrm{I}} + \de^{\mathrm{II}} + 
\de^{\mathrm{III}} } }
\int d^2\xa d^2\xb d^2\xc \frac{1}{|\xa - \xb|^{\ab\sqrt{\la}f}
|\xb - \xc|^{-\ab\ac\sqrt{\la}f}|\xc - \xa|^{\ac\sqrt{\la}f} }
\label{af}\end{equation}
with
\begin{equation}
f = \kappa^2 + \sum_{k=1}^2 r_k^2(\omega_k^2 + 
(m_k^{\mathrm{I}})^2 ) - \frac{2\kappa \de^{\mathrm{I}}}{\sqrt{\la}}.
\end{equation}
From the marginality condition of vertex operator the worldsheet
three-point correlator should take a 2d scaling behavior
$|\xa - \xb|^{-2}|\xb - \xc|^{-2}|\xc - \xa|^{-2}$ which in the 
large spin leads to
\begin{equation}
\kappa^2 + \sum_{k=1}^2 r_k^2(\omega_k^2 + (m_k^{\mathrm{I}})^2 )
- \frac{2\kappa \de^{\mathrm{I}}}{\sqrt{\la}} = 0.
\label{vir}\end{equation}
The circular $(J_1, J_2)$ string solution with winding numbers 
$(m_1, m_2)$ was constructed \cite{ART} from the off-diagonal Virasoro
constraint that agrees with (\ref{mj}) and the diagonal Virasoro 
constraint
\begin{equation}
\frac{2\kappa E}{\sqrt{\la}} - \kappa^2 = 2\sum_{k=1}^2
\sqrt{m_k^2 + \nu^2} \mathcal{J}_k - \nu^2 
\label{dia}\end{equation}
with $\mathcal{J}_k  = r_k^2\omega_k, \; \omega_k = \sqrt{m_k^2 + \nu^2}$.
The identity $\nu^2 = \sum_k r_k^2\nu^2 = \sum_k r_k^2 
(\omega_k^2 - m_k^2)$ shows that (\ref{vir}) coincides with (\ref{dia})
since dimension $\de^{\mathrm{I}}$ is
identified with the incoming string energy $E^{\mathrm{I}}= E$.

Because of the energy conservation (\ref{coe}), that is, 
$\de^{\mathrm{I}} = \de^{\mathrm{II}} + \de^{\mathrm{III}}$ the 
correlator (\ref{af}) exhibits the expected scaling behavior of the 
extremal three-point correlator
\begin{equation}
\frac{1}{(2a)^{2\de^{\mathrm{I}}}},
\end{equation}
where $2a$ is the distance between the vertex operator of the initial
string state I and the vertex operators of the final string states II, 
III in the boundary of the Euclidean $AdS_5$ space.

The energy-spin relation derived from (\ref{dia}) or (\ref{vir}) is
expressed in the $\la/(J^{\mathrm{I}})^2$ expansion with $J^I = \jaa + 
\jba$ as
\begin{equation}
\de^{\mathrm{I}} =  E^{\mathrm{I}} =  J^{\mathrm{I}} +  \frac{\la}
{2(J^{\mathrm{I}})^2}\sum_{k=1}^2 (m_k^{\mathrm{I}})^2J_k^{\mathrm{I}}
- \frac{\la^2}{8(J^{\mathrm{I}})^4}\sum_{k=1}^2 (m_k^{\mathrm{I}})^4
J_k^{\mathrm{I}} + \cdots.
\end{equation}
From this expression we use (\ref{eon}), (\ref{etw}) and (\ref{jon}),
(\ref{jtw}) to obtain
\begin{equation}
\de^{\mathrm{II}} =  J^{\mathrm{II}} +  \frac{\la\ab^2}
{2(J^{\mathrm{II}} )^2}\sum_{k=1}^2 (m_k^{\mathrm{I}})^2
J_k^{\mathrm{II}} - \frac{\la^2\ab^4}{8(J^{\mathrm{II}})^4}\sum_{k=1}^2
(m_k^{\mathrm{I}})^4J_k^{\mathrm{II}} + \cdots
\end{equation}
with $J^{\mathrm{II}} = \jab + \jbb$ and 
\begin{equation}
\de^{\mathrm{III}} =  J^{\mathrm{III}} +  \frac{\la\ac^2}
{2(J^{\mathrm{III}})^2}\sum_{k=1}^2 (m_k^{\mathrm{I}})^2
J_k^{\mathrm{III}} - \frac{\la^2\ac^4}{8(J^{\mathrm{III}})^4}
\sum_{k=1}^2 (m_k^{\mathrm{I}})^4J_k^{\mathrm{III}} + \cdots
\end{equation}
with $J^{\mathrm{III}} = \jac + \jbc$. Further, the relations (\ref{mtw}),
(\ref{mth}) yield the following expressions of the
energies of the two string fragments
\begin{eqnarray}
\de^{\mathrm{II}} &=&  J^{\mathrm{II}} +  \frac{\la}{2(J^{\mathrm{II}})^2}
\sum_{k=1}^2 (m_k^{\mathrm{II}})^2J_k^{\mathrm{II}} - \frac{\la^2}
{8(J^{\mathrm{II}})^4}\sum_{k=1}^2 
(m_k^{\mathrm{II}})^4J_k^{\mathrm{II}} + \cdots, \nonumber \\
\de^{\mathrm{III}} &=&  J^{\mathrm{III}} +  \frac{\la}
{2(J^{\mathrm{III}})^2}\sum_{k=1}^2 (m_k^{\mathrm{III}})^2
J_k^{\mathrm{III}} - \frac{\la^2}{8(J^{\mathrm{III}})^4}\sum_{k=1}^2 
(m_k^{\mathrm{III}})^4J_k^{\mathrm{III}} + \cdots.
\end{eqnarray}
These spin-energy relations for the outgoing strings II, III 
are accompanied with 
\begin{equation}
\mab \jab = \mbb \jbb, \hspace{1cm} \mac \jac = \mbc \jbc,
\end{equation}
which are also given from (\ref{mj}).

\section{Conclusion}

Using the integrated vertex operators we have computed a specific
three-point correlator of the heavy string vertex operators representing
the circular winding string states which are point-like in $AdS_5$ and 
rotating with two spins and two winding numbers in $S^5$, where two
locations of vertex operators are the same.

We have constructed a Schwarz-Christoffel mapping from the complex
plane with three punctures to the splitting string surface consisting
of three cylinders. Taking advantage of this special mapping 
and considering the appropriate saddle-point surface we
have semiclassically evaluated the correlator of the string vertex
operators located on three punctures. We have shown that the 
Euclidean continuation of the circular winding string solution
mapped on the complex plane with three punctures solves
the relevant equations of motion on the comlex plane with
the delta-function sources at three insertion worldsheet points
of vertex operators.

In the process of a simple separation of string 
such that the two outgoing string solutions are identical to the
incoming string solution, the worldsheet 
space position of splitting is 
determined by using the consistency condition for splitting and 
specified by the ratio of the relevant winding numbers.
The string splitting is characterized by the separation of 
the initial winding number into two winding numbers
at the self-interaction of the incoming
closed string so that winding numbers, energy and spins are conserved
in the same way.

By analyzing the restricted three-point correlator we have observed
that the marginality condition of the vertex operator for the 
worldsheet scaling behavior of the semiclassically evaluated 
three-point correlator yields the same dispersion relation among 
energy (dimension), spins and winding numbers as is obtained from
the diagonal Virasoro constraint for the circular winding string solution,
while the requirement for the worldsheet three-point correlator
 to have the 2d conformal invariant expression
gives the same relation between spins and winding numbers as follows
from the off-diagonal Virasoro constraint.

We have demonstrated that owing to the same separation behavior
in the conservations of energy, spins and winding numbers
the dispersion relations of the two outgoing strings become
the same expression as the dispersion relation of the initial
incoming string.

We have observed that the resulting three-point correlator 
shows the space-time scaling behavior of the extremal correlator
on the boundary theory, which is associated with the energy
conservation in the string splitting. This extremal behavior
seems to be related with the choice of the Schwarz-Christoffel
mapping which resembles the conformal mapping describing the
three-string interaction in the light-cone gauge for the 
string field theory.

\end{document}